\documentclass[
               twocolumn,
               prd,nofootinbib,
               showpacs,
               amsmath,amssymb]
              {revtex4}
\usepackage{graphicx,color}

\begin{document}

\title{Brane solution with an increasing warp factor}

\author{Michael~Maziashvili}
\email{maziashvili@ictsu.tsu.edu.ge} \affiliation{Department of
Theoretical Physics, Tbilisi State University, \\ 3 Chavchavadze
Ave., Tbilisi 0128, Georgia}

\begin{abstract}
  We present a new brane solution in the (1+5)-dimensional space-time with an increasing warp factor,
  which localizes the zero modes of all kinds of matter fields and
  Newtonian gravity. The interesting features of this solution
  are: 1) In contrast to the Gogberashvili-Singleton case the effective wave
  function of the zero mode spinor field confined on the
  brane is finite at a position of brane; 2) There is exactly one zero mode for each matter
  field and the gravitational field confined on the brane with a finite probability and infinitely
  many zero modes (labelled by the angular momentum corresponding to extra
  space) are located in the bulk at different localization radii.
\end{abstract}


\maketitle

\section{Introduction}

A new wave of activity in the field of extra dimensions came with
the framework of Antoniadis, Arkani-Hamed, Dimopoulos and Dvali
\cite{ADD} who observed that the Higgs mass hierarchy problem can
be addressed in models with large extra dimensions. The idea that
our world is a three brane embedded in a higher dimensional
space-time with non-factorizable warped geometry has been much
investigated since the appearance of papers \cite{Randall1,
Gogberashvili1}. In this idea, the key observation is that the
graviton, which is allowed to be free to propagate in the bulk, is
confined to the brane because of the warped geometry, thereby
implying that the gravitational law on the brane obeys the usual
four dimensional Newton's law as desired. A first particle physics
application of the brane world idea was put forward by Rubakov and
Shaposhnikov \cite{Rubakov} and independently by Akama
\cite{Akama}.

On the other hand, the other local fields except the gravitational
field are not always localized on the brane even in the warped
geometry. Indeed, in the Randall-Sundrum model in five dimensions
\cite{Randall1}, the following facts are well known: spin 0 field
is localized on a brane with positive tension which also localizes
the graviton while the spin 1/2 and 3/2 fields are localized not
on a brane with positive tension but on a brane with negative
tension \cite{Bajc}. Spin 1 field is not localized neither on a
brane with positive tension nor on a brane with negative tension
\cite{Pomarol}. In six space-time dimensions, the spin 1 gauge
field is also localized on the brane \cite{Oda2}.  Thus, in order
to fulfill the localization of Standard Model particles on a brane
with positive tension, it seems that some additional interactions
except the gravitational interaction must be also introduced in
the bulk. There is a lot of papers devoted to the different
localization mechanisms of the bulk fields in various brane world
models.

The brane solution with an increasing warp factor constructed by
Midodashvili, Gogberashvili and Singleton
\cite{Gogberashvili3,Gogberashvili2,Midodashvili} has a remarkable
feature that it can localize the zero modes of all kinds of matter
fields ranging from spin 0 scalar field to the spin 2
gravitational field by the gravitational interaction. To this
brane the effective wave function of the zero mode spinor field
that is confined on the brane becomes infinite at a position of
the brane. At first glance this singularity looks to be absolutely
harmless since it is integrable, but the main problem it causes is
related with the localization of a field theory model including
the interaction of matter fields with the fermion field in some
width of a brane.

In this paper we present a new brane solution to the Einstein
equations in (1+5)-dimensional space-time with an increasing warp
factor, which localizes the gravitational and matter fields with
the regular wave functions. In addition this brane localizes
infinitely many zero modes in the bulk at the different
localization radii from the brane. The brane solution is derived
in Sec. II. In Sec. III we consider the localization problem of
zero modes of matter and gravitational fields on the brane. Sec.
IV contains some concluding remarks.

\section{Brane solution}

In this section we closely follow the derivation of the brane
solution presented in papers \cite{Gogberashvili2}. The general
form of action of the gravitating system in six dimensions is
\begin{equation} \label{action}
S = \int d^6x\sqrt{- ^{(6)}g}\left(\frac{M^4}{2}\,^{(6)}R +
\Lambda +  {\cal L}\right),
\end{equation}
where $M$ is the fundamental scale, $^{(6)}R$ is the scalar
curvature, $\Lambda$ is the cosmological constant and ${\cal L}$
is the Lagrangian of matter fields. All of these quantities are
six dimensional. Throughout this paper the Greek indices refer to
the coordinates on the brane and take the range $0-3$, the lower
case Latin indices that correspond to the coordinates of a
transverse space take values $5,~6$, and the capital Latin indices
refer to the coordinates in the bulk $A,~B,\ldots=0-3,~5,~6$.

The 6-dimensional Einstein equations with stress-energy tensor
$T_{AB}$ are
\begin{equation} \label{Einstein6}
^{(6)}R_{AB} - \frac{1}{2} g_{AB}~^{(6)}R =
\frac{1}{M^4}\left(\Lambda g_{AB} + T_{AB}\right).
\end{equation}
The bulk space-time metric has the form
\begin{equation}\label{ansatzA}
ds^2  = \phi ^2 (r)\overline{g} _{\alpha \beta } (x^\nu )dx^\alpha
dx^\beta - \lambda (r)\delta_{ik}dx^idx^k,
\end{equation}
where the metric of ordinary 4-space, $\overline{g}_{\alpha \beta
}(x^\nu)$, has the signature $(+,-,-,-)$, $\delta_{ik}$ is the
Kronecker symbol, $r=\sqrt{\delta_{ik}x^ix^k}$, the summation
convention is used over repeated indices and all coordinates $x^A$
run in the interval $(-\infty,~\infty)$. So that the extra part of
eq.(\ref{ansatzA}) is a conformally flat 2-dimensional space.

The energy-momentum tensor is assumed to have the form
\begin{align} \label{source}
T_{\mu\nu} = - g_{\mu\nu} F(r), ~~ T_{\theta\theta} = r^2\lambda(r)K(r),\nonumber\\
T_{rr} = \lambda(r)K(r),~~T_{r\mu}=T_{\theta\mu}=0 ,
\end{align} where $r,~\theta$ are the polar coordinates in the $x^5,~x^6$ plane.
The energy-momentum conservation gives the following relation
between the source functions
\begin{equation} \label{deltaT}
K^{\prime} + 4\frac{\phi^\prime}{\phi} \left(K - F \right) = 0,
\end{equation} where the prime denotes differentiation with
respect to $r$. The equations (\ref{Einstein6}) are solved under
assumption that the 4-dimensional Einstein equations have the
ordinary form without a cosmological term
\begin{equation} \label{Einstein4}
R_{\mu\nu} - \frac{1}{2} \eta_{\mu\nu} R = 0.
\end{equation}
The Ricci tensor in four dimensions $R_{\alpha\beta}$ is
constructed from the 4-dimensional metric tensor
$\overline{g}_{\alpha\beta}(x^{\nu})$ in the standard way. Then
with the \emph{ans{\"a}tze} (\ref{ansatzA}) and (\ref{source}) the
Einstein field equations (\ref{Einstein6}) become
\begin{widetext}
\begin{align}\label{Einstein6a} 3 \frac{\phi^{\prime
\prime}}{\phi} + 3 \frac{\phi ^{\prime}}{r \phi} + 3
\frac{(\phi^{\prime})^2}{\phi ^2} + \frac{1}{2}\frac{\lambda
^{\prime \prime}}{\lambda }
-\frac{1}{2}\frac{(\lambda^{\prime})^2}{\lambda^2} +
\frac{1}{2}\frac{\lambda^{\prime}}{r\lambda } = \frac{\lambda
}{M^4}[F(r) - \Lambda ] , \nonumber & \\\frac{\phi^{\prime}
\lambda^{\prime}}{\phi \lambda } + 2 \frac{\phi^{\prime}}{r\phi} +
3 \frac{(\phi^{\prime})^2}{\phi ^2} = \frac{\lambda }{2 M^4}[K(r)
- \Lambda ], \\2\frac{\phi^{\prime \prime}}{\phi} -
\frac{\phi^{\prime} \lambda ^{\prime}}{\phi \lambda } +
3\frac{(\phi^{\prime})^2}{\phi^2} = \frac{\lambda }{2 M^4}[K(r) -
\Lambda ].\nonumber
\end{align}

\end{widetext}
These equations are for the $\alpha \alpha$, $rr$, and $\theta
\theta$ components respectively. Subtracting the $rr$ from the
$\theta \theta$ equation and multiplying by $\phi / \phi
^{\prime}$ we arrive at
\[
\label{phi-g} \frac{\phi^{\prime \prime}}{\phi^{\prime}} -
\frac{\lambda^{\prime}}{\lambda } -\frac{1}{r} = 0 .
\]
This equation has the solution
\begin{equation}
\label{g} \lambda (r)= \frac{\rho^2 \phi ^{\prime}}{r} ,
\end{equation}
where $\rho$ is an integration constant with units of length. By
taking into account the eq.(\ref{g}), the system of equations
(\ref{Einstein6a}) reduces to one independent equation. Taking
either the $rr$, or $\theta \theta$ component of these equations
and multiplying it by $r\phi^4$ gives
\begin{equation}\label{rr}
r \phi^3 \phi^{\prime \prime} + \phi^3 \phi^{\prime} + 3r\phi^2
(\phi^{\prime})^2 = \frac{\rho^2 \phi^4 \phi^{\prime}}{2 M^4}
[K(r) - \Lambda ].
\end{equation} Notice that the left-hand side of eq.(\ref{rr}) is equal to $(r\phi^3\phi')'$. The eq.(\ref{Einstein6a}) admits the solution of the form
\begin{equation}\label{newbrane}\phi=1+a\tanh(r^2/\epsilon^2),~~\lambda=\frac{2a\rho^2}{\epsilon^2\cosh^2(r^2/\epsilon^2)},\end{equation}
where $\epsilon$ is an integration constant with units of length,
which corresponds to the following rather complicated source
functions expressed in term of $\phi$ as
\begin{widetext}
\begin{equation}\label{ugsorfunc}K(\phi)=\Lambda+\frac{4M^4}{\rho^2\phi}\left\{1-\frac{1}{2}\ln\left(\frac{a+\phi-1}{a+1-\phi}\right)\left[2\frac{\phi-1}{a}+\frac{3a}{\phi}\left(1-\frac{(\phi-1)^2}{a^2}\right)\right]\right\},~~F(\phi)=K(\phi)+\frac{\phi}{4}\frac{dK}{d\phi},\end{equation}
\end{widetext} where $a$ is an arbitrary dimensionless positive parameter. As it is seen from
eq.(\ref{newbrane}), the warp factor $\phi$ monotonically
increases from $1$ to $1+a$ as $r$ goes from $0$ to $\infty$. The
brane (\ref{newbrane}) is located at $x^5=x^6=0$ and its width is
characterized by $\epsilon$. It is straightforward to generalize
this solution to the case when the bulk space-time dimension is
grater than $6$ and the brane codimension $\geq 2$ \cite{Oda3}.

The brane solution constructed in
\cite{Gogberashvili3,Gogberashvili2,Midodashvili}, which will be
refereed to as a GS brane in what follows, corresponds to the
following choice of the source functions
\[K(\phi)=\frac{c_1}{\phi^2}+\frac{c_2}{\phi},~~F(\phi)=\frac{c_1}{2\phi^2}+\frac{3c_2}{4\phi},\]
where $c_1,~c_2$ are constants, and has the form
\begin{equation}\label{GSM}\phi=\frac{3\epsilon^2+br^2}{3\epsilon^2+r^2},~~\lambda=\frac{9\epsilon^4}{(3\epsilon^2+r^2)^2},\end{equation}
where constant $b>1$. The extension of this solution to the case
when the codimension of the brane is greater than $2$ is given in
\cite{Singl}.

\section{Localization of matter and gravitational fields}

Now let us consider the localization problem of different kinds of
matter fields on the brane (\ref{newbrane}). We treat the
dependence of $\sqrt{- ^{(6)}g}\,{\cal L}$ on the extra dimensions
as (effective) wave functions for the Kaluza-Klein modes.

{\it \bf a) Scalar field.} The transverse equation for the scalar
field $\Phi(x^{\mu})\varphi(r)\exp(in\theta)$, where $n$ is an
arbitrary integer number, reads
\begin{equation}\label{zeroscaln}\varphi''+\left(\frac{1}{r}+\frac{4\phi'}{\phi}\right)\varphi'-\frac{n^2}{r^2}\varphi =0.\end{equation}
The wave function of scalar $n$'th zero mode has the form
\begin{equation}\label{scalwave}\triangle^{(0)}_n(x^5,~x^6)\propto\varphi^2\phi^2\lambda.\end{equation}
The zero mode field with zero angular momentum
$n=0,~\varphi=const.$ is located on the brane with the probability
$\triangle^{(0)}_0(0,~0)\propto a\rho^2/\epsilon^2$ and the
localization width$\sim\epsilon$. The second solution of
eq.(\ref{zeroscaln}) with the zero angular momentum for
$r\rightarrow 0,~\infty$ behaves as $\varphi\sim\ln(r/c)$, where
$c$ is integration constant with units of length. This solution
gives the normalizable wave function too. Taking $c\ll\epsilon$
one concludes that since for this solution $\varphi=0$ at $r=c$
this zero mode is located either on the brane (with an infinite
probability) or in the region $r>c$. The regular solutions at
$r=0$ of eq.(\ref{zeroscaln}) for $n\neq 0$ are localized in the
bulk at some radii from the brane. Namely, by taking into account
the eq.(\ref{newbrane}), one finds that the regular solutions of
eq.(\ref{zeroscaln}) at the origin ($r=0$) behave as $r^{|n|}$ for
$r\rightarrow 0,~\infty$. As it is clear from eq.(\ref{scalwave})
these solutions give the normalizable wave functions. The greater
$|n|$, the larger the localization radius. The GS brane
(\ref{GSM}) localizes only one, $n=0$, mode exactly on the brane.
The corresponding wave function obtained from eq.(\ref{GSM}) has
the form
\[\triangle^{(0)}_{GS}(x^5,~x^6)\propto\frac{\left(3\epsilon^2+br^2\right)^2}{\left(3\epsilon^2+r^2\right)^4}.\]

{\it \bf b) Spinor field.} Now we turn to the spinor field
localization problem on the brane (\ref{newbrane}). Due to Fock's
formalism \cite{Fock}, for constructing the Dirac operator in the
gravitational background one has to introduce the local
Minkowskian frame at each point of space.\footnote{For the
diagonal metric there is a simple prescription to construct the
Dirac operator immediately \cite{Fock}.} Introducing the following
\emph{sechsbein}
\begin{equation}e_A^{~~B}=\left(\delta_{\mu}^B\phi^{-1},~\delta_5^B\lambda^{-1/2},~\delta_6^B\lambda^{-1/2}\right),\end{equation}
where the first index corresponds to the flat tangent 6d Minkowski
space, the transverse Dirac operator takes the form
\begin{equation}D=\Gamma^k\left\{\partial_k+\frac{\delta_{km}x^m}{r}\left(2\frac{\phi'}{\phi}+\frac{\partial_r\sqrt{\lambda}}{2\sqrt{\lambda}}\right)\right\}.\end{equation}
The Dirac $\Gamma$-matrices used in the paper are given in
Appendix. Taking the zero mode spinor field to be
\begin{equation}\label{spinzero}\Psi=\left(\begin{array}{ll}\Psi_1(x^{\mu})\psi_1(r)\\
\Psi_2(x^{\mu})\psi_2(r)
\\\Psi_3(x^{\mu})\psi_3(r)\\\Psi_4(x^{\mu})\psi_4(r)
\end{array}\right)e^{in\theta},\end{equation} where $\Psi_1,~\Psi_2,~\Psi_3,~\Psi_4$ are two-component spinors, the equation for zero mode
$D\Psi=0$ reduces to
\[\begin{array}{ll}
\left(\partial_r+2\phi'/\phi+\partial_r\sqrt{\lambda}/2\sqrt{\lambda}+n/r\right)\psi_{1,3}=0,\\\\
\left(\partial_r+2\phi'/\phi+\partial_r\sqrt{\lambda}/2\sqrt{\lambda}-n/r\right)\psi_{2,4}=0.
\end{array}\] Thus
\[\psi_{1,3}\propto\phi^{-2}\lambda^{-1/4}r^{-n},~~
\psi_{2,4}\propto\phi^{-2}\lambda^{-1/4}r^{n}.\] The wave function
for the $n$'th zero mode of $\Psi_{2,4}$ spinor fields takes the
form
\begin{equation}\label{spin24}\triangle^{(1/2)}_n(x^5,~x^6)\propto\phi^{-1}\lambda^{1/2}r^{2n}.\end{equation}
Analogously, for the wave function of $n$'th zero mode of
$\Psi_{1,3}$ spinor fields one finds
\begin{equation}\label{spin13}\triangle^{(1/2)}_n(x^5,~x^6)\propto\phi^{-1}\lambda^{1/2}r^{-2n}.\end{equation}
For $n=0$ the fields $\Psi_{1,3}=\Psi_{2,4}$ are localized on the
brane with probability
$\triangle^{(1/2)}_0(0,~0)\propto\sqrt{a}\rho/\epsilon$. As it is
seen from eqs.(\ref{spin24},~\ref{spin13}) the fields
$\Psi_{2,4}\left(\Psi_{1,3}\right)$ for $n>0~(n<0)$ are localized
in the bulk. The GS brane localizes the spinor zero mode provided
$n$ is a half-integer number with the wave function
\begin{equation}\label{GSspinorwave}\triangle^{(1/2)}_{GS}(x^5,~x^6)\propto\frac{1}{r\left(3\epsilon^2+br^2\right)}.\end{equation}

{\it \bf c) Vector field.} As a next step we consider spin $1$
gauge field. The transverse equation for the $U(1)$ vector field
$A_{\nu}(x^{\mu})\chi(r)\exp(in\theta),~A_5=A_6=0$ takes the form
\begin{equation}\label{veczermodeeq}\chi''+\left(\frac{1}{r}+4\frac{\phi'}{\phi}-\frac{\lambda'}{\lambda}\right)\chi'-\frac{n^2}{r^2}\chi=0.\end{equation}
For the zero mode wave function one obtains
\begin{equation}\label{vecwave}\triangle^{(1)}_n(x^5,~x^6)\propto\chi^2\lambda.\end{equation}
The $n=0$ mode $\chi=const.$ is located on the brane with a finite
probability. The second solution, which for $r\rightarrow 0$
behaves as $\chi\sim\ln(r/c)$ and for $r\rightarrow\infty$ as
$\chi\sim\int_0^r\exp(-2x^2/\epsilon^2)dx$, under assumption
$c\ll\epsilon$ is located either on the brane (with an infinite
probability) or in the region $r>c$. For $n\neq 0$ the asymptotic
behavior of $\chi$ that corresponds to the convergent wave
function (\ref{vecwave}) at the origin ($r=0$) is given by
$r^{|n|}$ when $r\rightarrow 0$. To find the asymptotic behavior
when $r\rightarrow\infty$ we transform the eq.(\ref{veczermodeeq})
by the transformation $\chi=vr^{-1/2}\exp(-r^2/\epsilon^2)$ to the
form
\[v''-\left[\frac{4}{\epsilon^2}+\left(n^2-\frac{1}{4}\right)r^{-1}+\frac{4}{\epsilon^4}r^2\right]v=0,\]
and use the WKB approximation which is valid to this equation for
large values of $r$, see \cite{MW}. The asymptotic solutions
obtained in this way read \[\chi\sim
r^{-1}\exp(-r^2/\epsilon^2)\exp(\pm r^2/\epsilon^2).\] Both these
solutions give the normalizable wave function (\ref{vecwave}) for
$r\rightarrow\infty$. Thus, the zero mode vector fields $n\neq 0$
are located in the bulk. For the zero mode $n=0$ wave function in
the case of GS brane one obtains
\[\triangle^{(1)}_{GS}(x^5,~x^6)\propto\frac{1}{\left(3\epsilon^2+r^2\right)^2}.\]

{\it \bf d) Gravitational field.} For the graviton fluctuations
\[g_{\mu\nu}=\phi^2(r)\left\{\eta_{\mu\nu}+h_{\mu\nu}(x^{\alpha})\sigma(r)\exp(in\theta)\right\},\]
where $\eta_{\mu\nu}=$diag$(+,-,-,-)$, the transverse equation for
the zero modes is analogous to the eq.(\ref{zeroscaln}) with
$\varphi$ replaced by $\sigma$ \cite{Gogberashvili2}. Thus the
localization property of the gravitational field is the same as
for the scalar field.

\section{Conclusions}

A new brane solution in the (1+5)-dimensional space-time with an
increasing warp factor is presented. This brane localizes the zero
modes of all kinds of matter fields and Newtonian gravity by the
gravitational interaction. Besides the zero mode that is located
on the brane with a finite probability, there is an additional
zero mode with zero angular momentum, which behaves as
$\sim\ln(r/c)$ in the vicinity of $r=0$. But one needs not to
worry about this mode because the wave function corresponding to
this solution becomes zero at $r=c$ (we assume that
$c\ll\epsilon$) and one can merely assume that this mode is
confined in the region $r>c$. The zero modes $n\neq 0$ are located
in the bulk at different radii from the brane. The greater $|n|$
the larger the localization radius.  The fact that for
$n=-1,~0,~1$ the brane localizes two two-component spinors of the
opposite chirality in the width of a brane may be interesting for
explanation of the fermionic family replication and the fermionic
mass hierarchy. For instance, localization of different species of
fermions at different points of a thick brane was used to solve
the hierarchy problem in the split fermion model \cite{Split}.
Notice that in the case of GS brane all matter fields are located
exactly on the brane and the different localization radii for the
different kinds of matter fields indicated in
\cite{Gogberashvili2} are due to fact that the radial variable $r$
coming in polar coordinates from the integration measure is
included in the wave function \cite{Note}. The fact that the wave
function (\ref{GSspinorwave}) becomes infinity at a position of
brane prevents the localization of the field theory model
including the interaction of matter fields with the fermion field
in some neighborhood of the brane. For the presented brane
solution this problem is absent because the fermion wave function
is finite on the brane. We note that the localization of fermion
field on the GS brane requires the fermion field to satisfy the
antiperiodic boundary condition with respect to $\theta$. In the
present approach, the presence of a solution to Einstein's
equations heavily depends on the form of the source functions. For
obtaining of the present brane solution we have used the ugly
source functions (\ref{ugsorfunc}). It is desirable to construct
the source functions from fundamental matter fields such that to
insure the stability of the brane.

\begin{acknowledgments}
The author is greatly indebted to M.~Gogberashvili for many
stimulating discussions, for reading this manuscript and useful
comments. The author is also indebted to G.~Skhirtladze and
L.~Tskipuri for help during the work on this paper.
\end{acknowledgments}
\vspace{0.4cm}
\section*{Appendix}

\subsection*{$\Gamma$ matrices}
The chiral representation of the 6d flat Dirac $\Gamma$-matrices
we use in the paper is
\[
\Gamma^{\mu}=\left(\begin{array}{cc}
0 & \gamma^0\gamma^{\mu}\\
\eta_{\mu\nu}\gamma^0\gamma^{\nu} & 0
\end{array}\right),\ \
\Gamma^5 =\left(\begin{array}{cc}
0 & i\gamma^0\gamma^5\\
-i\gamma^0\gamma^5 & 0
\end{array}\right),\]
\[
\Gamma^6=\left(\begin{array}{cc}
0 & \gamma^0\\
\gamma^0 & 0
\end{array}\right), \ \
\gamma^{\mu}=\left(\begin{array}{cc}
0 & \sigma^{\mu}\\
\eta_{\mu\nu}\sigma^{\nu} & 0
\end{array}\right),\]
where $\gamma^5=$diag$(1,~-1),~\sigma^0$ is unit matrix and
$\sigma^1,~\sigma^2,~\sigma^3$ are the Pauli matrices. For the
eight component spinor (\ref{spinzero}) one simply finds that
spinors $\Psi_{1,4}$ and $\Psi_{2,3}$ have the opposite chirality
from the four dimensional point of view.

\end{document}